\documentclass[11pt]{article}
\begin{document}
    \begin{center}
           {\large\bf Anomalous Magnetic Moment of 
                     the Muon in a Composite Model  
           } \\ \bigskip\bigskip
           {\large\sl Prasanta Das\footnote{E-mail: 
                pdas@mri.ernet.in}
           },
           {\large\sl Santosh Kumar Rai\footnote{E-mail: 
                skrai@iitk.ac.in}
           },
           {\large\sl Sreerup Raychaudhuri\footnote{E-mail: 
                sreerup@iitk.ac.in}
           } \\
           {\rm Department of Physics, 
                Indian Institute of Technology, 
                Kanpur 208 016, India.}
          \\ ({\it March 31, 2004} )
    \end{center}
\vspace*{0.2cm}
    \begin{noindent}
    \small
           We make a fresh evaluation of contributions to ($g$-2) of the
           muon in the framework of a preonic model with coloured
           vector-like leptons and heavy coloured $Z$ bosons and discuss
           their implications in the light of the recent Brookhaven
           measurement. It is shown that the observed deviation(s) can be
           accommodated within this framework with masses and couplings
           consistent with all constraints.
    \normalsize
    \end{noindent}
\vspace*{0.5cm}

Ever since the original QED derivation of Schwinger\cite{Schwinger}, the
anomalous magnetic moment of the electron --- and the muon --- has been a
testing ground for the correctness of the Standard Model (SM).  The most
precise measurement of this parameter has been the recent
measurement\cite{BNLmuon}
\begin{equation}
    a_\mu = \frac{g - 2}{2} = 11~659~208~(6)~\times~10^{-10}
\label{expt}
\end{equation}
by the E821 experiment at the Brookhaven National Laboratory (BNL), which
is the culmination of a series of measurements of ever-increasing accuracy
over the past several years. Given the care and effort which has gone into
this measurement, it is probably fair to say that this result is one of the
most accurate ever achieved and a triumph of human ingenuity.

The theoretical situation is somewhat more murky. In the SM, the anomalous 
magnetic moment of the muon arises from electromagnetic, weak and hadronic 
sources.  While there is no problem in determining the first two using the 
basic electroweak theory, it is not possible to make a perturbative estimate 
of the hadronic contributions because the strong coupling constant is too 
large at the muon scale. Accordingly the lowest order vacuum polarisation 
in the hadronic contribution is calculated using a dispersion integral and 
the actual data on $e^+e^-$ annihilation to hadrons. With this input, the 
theoretical prediction is 
\begin{equation}
    a_\mu^{(e^+e^-)} = 11~659~181~(8)~\times~10^{-10}
\end{equation}
which means that the present measurement (Eqn.~\ref{expt}) deviates
from the SM by
\begin{equation}
a_\mu^{\rm exp} - a_\mu^{(e^+e^-)} = (2.7~\pm~1.0)~\times~10^{-9}
\end{equation}
which represents a 2.7$\sigma$ excess. On the other hand, the hadronic 
vacuum polarisation could also be determined by using data from hadronic
$\tau^\pm$ decays in combination with the CVC hypothesis, and this yields
\begin{equation}
    a_\mu^{(\tau^+\tau^-)} = 11~659~196~(8)~\times~10^{-10}
    \end{equation}
which is just a little more than 1$\sigma$ away from the experimental result.

In the normal course of events, one would trust the calculation of $a_\mu^{(e^+e^-)}$
much more than that of $a_\mu^{(\tau^+\tau^-)}$, because the $e^+e^-$ data is
much more clean and the number of theoretical assumptions are less. If we accept 
this, then it appears that the E821 experiment has observed the first
real deviation from the SM in experimental data, which makes their result
very exciting. This deviation constitutes an immediate challenge to
theorists to find possible explanations, and several such analyses have
already appeared in the literature. However, since scientists tend to be 
conservative, the $a_\mu^{(\tau^+\tau^-)}$ calculation cannot and should not 
be ignored --- especially as it is in much better agreement with the SM, which
has been tested with spectacular success in so many other areas. 

In this article,
we shall take the middle ground, {\it viz.} we shall try to provide an explanation of
the nearly 3$\sigma$ discrepancy between the measured $a_\mu^{\rm exp}$ and the
calculated $a_\mu^{(e^+e^-)}$ in a model with new physics, but keeping in mind the 
fact that accepting the $a_\mu^{(\tau^+\tau^-)}$ calculation would provide strong
constraints on any such model.
It has been pointed out\cite{Marciano} that in order to explain a
discrepancy of the observed magnitude in a low-energy phenomenon like the
muon anomalous magnetic moment, it is necessary to invoke new physics with
particle masses around the electroweak scale as well as couplings of at
least electroweak strength. This makes it immediately clear that the new
physics responsible for the BNL anomaly cannot be due\cite{Kane} to
effects expected at ${\cal O}{\rm (1~TeV)}$, such as, for example,
graviton exchanges in theories with extra dimensions\cite{Graesser} and
possible non-commutative effects in QED\cite{Jabbari}, which are two of
the currently fashionable ideas. On the other hand, supersymmetry, which
is the most popular way of going beyond the SM, can explain the observed
excess, since it predicts new particles with the correct order of masses
and couplings. In fact, several studies have already appeared in the
literature\cite{Kane,SUSY} which map out the portion of the supersymmetric
parameter space which could give rise to the observed effect.

While it is gratifying to know that supersymmetric models constitute an
acceptable explanation of the muon anomaly, it is also relevant and
interesting to ask what {\it other} new physics options can give rise to such an
effect. Suggestions already made in the literature vary from a scalar
leptoquark exchange\cite{Mahanta,Choudhury} to lepton
flavour-violation\cite{Huang}, exotic vector-like fermions\cite{Rakshit},
torsion fields in non-Einsteinian gravity models\cite{Uma}
and possible non-perturbative effects at the 1 TeV
order\cite{Lane,Mahanta2}. Given the importance of the BNL result, many
more suggestions will surely be forthcoming, in the fullness of time.

In this work, we revive earlier ideas\cite{composite} that the excess
contribution to the muon anomaly could be a signal for compositeness of
the muon and of the weak gauge bosons. Given the well-known history of
finding layers of substructure every time a significant increase in the
resolving power of experiments has increased, it is surely reasonable to
address the question whether the BNL anomaly could be the first hint of a
new layer of substructure which may be discovered some time in the future.
However, it is also a well-known fact that since the original 
suggestion\cite{PatiSalam}
several -- in fact, one may say, dozens --- of preonic models have been
proposed in the literature. Different models among these have different
virtues and various motivations, and it is not our purpose to make a
comprehensive survey of these models\cite{Lyons}. A generic feature of all
these models, however, is the existence of {\it excited} leptons and gauge
bosons, which are just excited states of the preonic combinations which
make up ordinary leptons and gauge bosons. These excitations may be simple
orbital excitations, which share the same set of flavour and colour
quantum numbers as the SM particles, or they may have exotic flavour and
colour charges. To fix our ideas, we concentrate on a well-known model,
namely the so-called {\it haplon} model of Fritzsch and
Mandelbaum\cite{Fritzsch}, which has the prime virtues of simplicity and
elegance\footnote{Another popular model is the {\it rishon} model of
Harari and Seiberg\cite{Harari}.  This has a simpler preonic spectrum than
the haplon model, but by virtue of combining together several preons at a
time it predicts many more exotic colour states.}.

The haplon model is based on the assumption that the leptons, quarks and
weak gauge bosons of the SM (as also possible neutral and charged scalars)
are not elementary particles, but are composed of pairs of preons called
{\it haplons}. The fundamental symmetry of Nature is $SU(3)_c \times
U(1)_{em} \times SU(N)_h$, which, of course, makes the photon and the
gluons fundamental particles. The weak interaction is no longer a gauge
interaction, but is interpreted in this scenario as a feeble van der
Waals-type effect of the preonic gauge interaction $SU(N)_h$ (A variant
with just a $U(1)_h$ instead of $SU(N)_h$ has also been proposed; however,
it is hard to see how this could lead to preon confinement.) In this
scenario, the weak isospin and hypercharge have a status similar to the
isospin and hypercharge of the meson and baryon multiplets. The haplon
model is, in many ways, a copy of the quark model at a deeper level, with
$SU(3)_c \times U(1)_{em}$ replacing $U(1)_{em}$ and $SU(N)_h$ for the
preonic interactions replacing $SU(3)_c$ for the quark interactions. The
basic building blocks of matter, then, apart from the photon and gluon,
are two fermionic preons $\alpha({\bf 3}, -\frac{1}{2}, {\bf N})$ and
$\beta({\bf 3}, \frac{1}{2}, {\bf N})$, and two scalar preons $x({\bf 3},
-\frac{1}{6}, \bar{\bf N})$ and $y(\bar{\bf 3}, \frac{1}{2}, \bar{\bf
N})$. Using these, we can build up the entire SM by taking bound states of
pairs of haplons, and interpret the three observed fermion families as
representing orbital excitations of these bound states. Since it is known
that sequential or mirror fermions other that these three families are
ruled out by precision electroweak data\cite{PDG}, any further families
must be vector-like in nature. The cause for these being vector-like,
whereas the first three states are sequential, must be attributed to the
dynamics of the bound state, which are as yet unknown.

A natural consequence of having bound pairs of particles belonging to
either {\bf 3} or $\bar{\bf 3}$ of $SU(3)_c$, is to predict the existence
of exotic particles such as
\begin{itemize}
\item Colour-octet $W^\pm$ bosons ($W^+_8 = \bar{\alpha} \beta$ in
a spin-1 state);
\item Colour-octet $Z$ bosons 
($Z_8 = \frac{1}{\sqrt{2}}(\alpha\bar{\alpha} + \beta\bar{\beta})$ in a
spin-1 state);
\item Colour-octet neutral scalars ($H_8 = 
\frac{1}{2}(\alpha\bar{\alpha} + \beta\bar{\beta} + x\bar{x} + y\bar{y}$)
in a spin-0 state);
\item Colour singlet and octet charged scalars ($H^+_8 = 
\bar{\alpha} \beta$ in a spin-0 state);
\item Colour singlet and octet leptoquarks ($\Phi^{+2/3} = \bar{x}y$);
\item Colour sextet quarks ($u_6 = \bar{\alpha}\bar{x}$, 
$d_6 = \bar{\beta}\bar{x}$);
\item Colour-octet leptons ($\ell^-_8 = \bar{\beta}\bar{y}$, 
$\nu_{\ell 8} = \alpha y$).
\end{itemize}

Since such exotic coloured objects would be produced copiously at hadron
colliders, it is clear from their non-observation that that they must be
rather heavy. In fact, direct searches for the coloured fermions leads to
the following bounds\cite{PDG}: $m_{q_6} > 84$ GeV, $m_{\ell_8} > 86$ GeV
and $m_{\nu_8} > 110$ GeV. It is also clear from the non-observation of
deviations from the SM predictions at hadron colliders that coloured $W_8$
and $Z_8$ bosons, as well as scalar leptoquarks (coloured or otherwise)
must be rather heavy, with masses typically of the order of a few hundred
GeV or a few TeV. Direct detection of such objects --- if they exist ---
may, therefore, have to await the commissioning of the LHC. However, the
$a_\mu$ measurement at BNL may just provide a window into this mass regime
through virtual effects, and this is the motivation of the present work.

We address the question as to whether the existence of so many exotic
particles can affect the anomalous magnetic moment of the
muon\cite{composite}. It may be noted that the current deviation from the
SM is typically an effect of the order of $10^{-9}$, which makes it
comparable to relativistic effects in everyday life. It is interesting to
speculate that just as the tiny effect observed (or not observed) by 
Michelson was the first harbinger of relativity, so the tiny discrepancy 
observed at BNL could be the first
indication of substructure. To this end we have considered the possible
contributions to an excess muon magnetic moment from one-loop diagrams
involving the {\it exotic} particle spectrum. Of these, the ones which are
of greatest interest are of the following three types:
\begin{enumerate}
\item Vertex-type diagrams with a $W^\pm_8$ and/or a $H^\pm$/$H_8^\pm$
together with a neutrino (coloured or ordinary, as the case may be) in the
loop;
\item Vertex-type diagrams with a $Z^0_8$ and/or a $H^0$/$H_8^0$ 
together with a muon (coloured or ordinary, as the case may be) in the
loop;
\item Vertex-type diagrams with a scalar leptoquark $\Phi^{\pm 2/3}$ or
$\Phi_8^{\pm 2/3}$ and a quark (ordinary or colour sextet) in the loop.
\end{enumerate}
These follow the topology illustrated in Fig.~1($ii$).

In this work, we assume that the interactions of scalar bosons, coloured
or otherwise, mimic those of the SM Higgs bosons, {\it i.e.} these scalars
couple to particles with a strength proportional to the particle mass.  As
a result, their coupling to muons is very feeble and hence the effects of
these scalars need not be considered any further. The situation is
different for scalar leptoquarks, whose couplings have no SM analogy.
While there are strong restrictions on the masses and couplings of scalar
leptoquarks from the Fermilab Tevatron data\cite{PDG}, it has nevertheless
been shown\cite{Mahanta,Choudhury} that scalar leptoquark exchanges can
produce the desired excess in $(g - 2)$ of the muon. In this work,
however, we exclude this interesting possibility. We also assume that the
$SU(N)_h$ scale is very high ---  too high for any appreciable magnetic
moment to arise {\it directly} from orbital excitations of the preons\cite{Lane}.
Both of these possibilities have already been explored in the literature.
We are thus concentrating on the remaining one out of three different ways in 
which an excess muon magnetic moment can be generated in a composite model.

The detailed Feynman rules for the vector boson interactions in the haplon
model are given in Ref.~\cite{Gounaris}. An interesting feature of these
is that the interactions of the $W_8^\pm$ are chiral, whereas those of the
$Z_8$ are vector-like. It is also worth noting that there is no mixing
between the $Z_8$ and the photon, for obvious reasons, and hence, there is no
analogue of the Weinberg angle in $Z_8$ interactions. The fact that the
$Z_8$ interactions are vector-like allows us to evade any constraints from
the electroweak precision data, provided we allow the colour octet muon to
be practically degenerate with its coloured neutrino partner\cite{PDG}.

\begin{figure}[htb]
\begin{center}
\vspace*{2.4in}
      \relax\noindent\hskip -4.4in\relax{\includegraphics{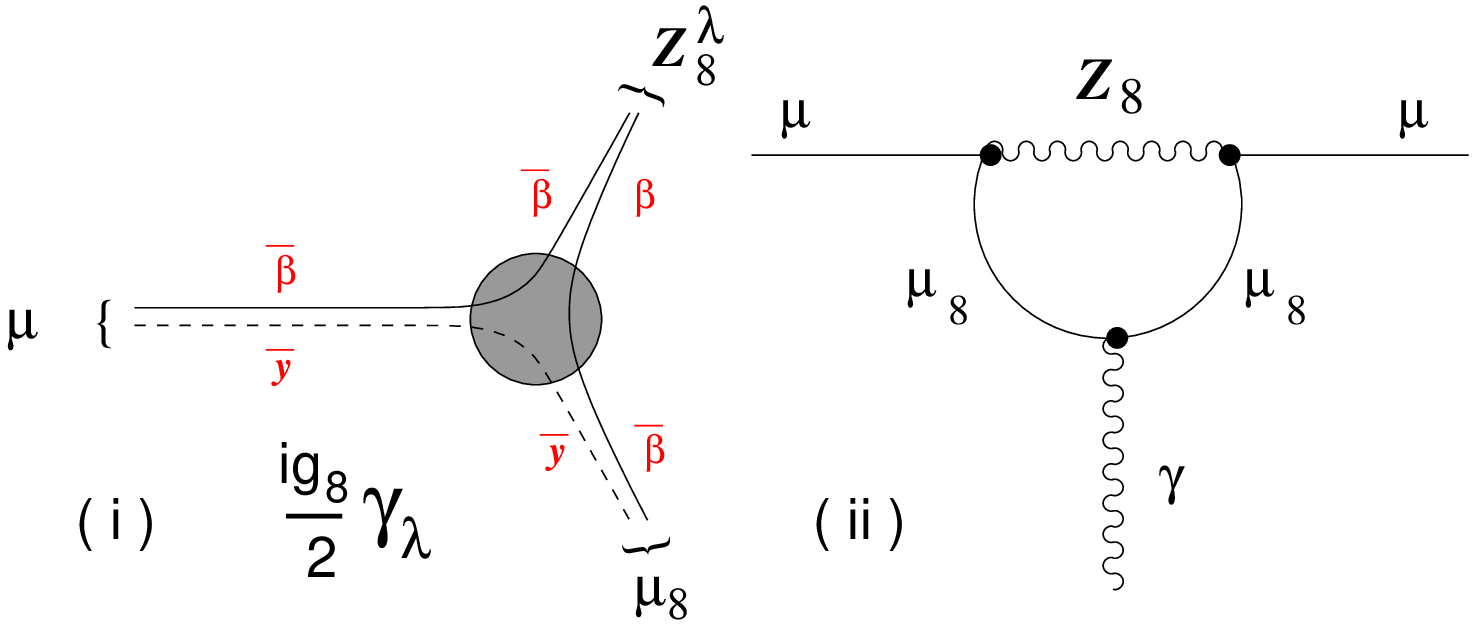}}
\end{center}
\end{figure}
\noindent {\bf Figure 1}.
{\footnotesize\it Illustrating ($i$)~the preonic level diagram responsible
for the $\mu$-$\mu_8$-$Z_8$ vertex $\frac{ig_8}{2}\gamma_\lambda$, and
($ii$)~the corresponding one-loop diagram contributing to $(g - 2)$ of the
muon. } 
\vskip 5pt

It is well-known\cite{Marciano} that chiral interactions lead to
contributions to the muon anomalous magnetic moment which are strongly
suppressed by the ratio of muon mass to the (large) scale of the new
physics, and hence, it in not useful to consider the first type of diagram
listed above. On the other hand, it is indeed possible to evaluate the new
physics contribution to the anomalous magnetic moment of the muon at the
one-loop level using the vector-like $Z_8$-$\mu^\pm_8$-$\mu^\mp$
interaction illustrated in Fig.~1. It is worth noting that the coupling
constant $g_8$ --- again arising from preon dynamics --- is unknown but is
expected to be of electroweak strength, just like the well-known
$Z$-$\mu^+$-$\mu^-$ coupling.  We have two more unknown quantities to
consider in Fig.~1, viz. the mass $m_8$ of the octet muon ($\mu_8$), and
the mass $M_8$ of the octet $Z$-boson ($Z_8$). These are expected to be at
least a few hundred GeV and may well be much larger.

At this point, it is relevant to ask if there are analogous contributions
to the self-energy of the muon and the answer is, obviously, yes. In fact,
the muon mass receives large contributions from all of the above 
Green's functions, if one simply removes the photon leg. Consequently, the
haplon model, like many others of its kind, must be severely constrained by the
experimental data on the muon mass\cite{Marciano}. There are two
possibilities:
\begin{enumerate}
\item The coupling of colour octet $W_8$ and $Z_8$ (and leptoquark) bosons
to the muon and its colour octet conterpart could be very small 
($< 10^{-4}$), and consequently, all contributions to the self-energy would
be severely suppressed.
\item The relevant couplings could be large ($\sim 0.1$), but the (large)
contributions to the muon self-energy from diagrams with $Z_8$-$\mu_8$
exchange and $W_8$-$\nu_8$ exchange could cancel to give the correct order
of magnitude. Phenomenologically, this can always be arranged by tuning
the masses and couplings of the colour octet particles suitably.
\end{enumerate}
It is important to note that {\it both} these cases involve some fine-tuning.
There is no \'a priori reason to assume that van der Waals-type
interactions will be suppressed when some of the particles are colour
octets --- which makes small ($\sim 10^{-4}$) couplings unnatural; at the
same time, there is no symmetry which guarantees the cancellation of
self-energy contributions of different diagrams to give a net contribution
two to three orders smaller than either contribution. However, one of the
two mechanisms {\it must} be at work, since, after all, the muon mass
measurements do not allow large self-energy contributions. Of course,
the reason must
ultimately be sought in the non-perturbative preon dynamics. As these are,
at present, unknown, we must work within a set of phenomenological assumptions.

In the first of the above cases, of course, the colour octet bosons and fermions 
are essentially decoupled from the SM fermions and hence, contributions to 
the muon anomalous magnetic moment are as severely suppressed as the self-energy.
The haplon model cannot then constitute an explanation for the BNL anomaly 
and hence, the story ends at this point. However, the second option is more
interesting, since the chiral structure of the $W_8^\pm$-$\nu_8$-$\mu^\mp$
and $Z_8$-$\mu_8^\pm$-$\mu^\mp$ couplings guarantee that there will be
near-vanishing contributions to the muon anomalous magnetic moment from
diagrams with $W_8$-$\nu_8$ loops. Thus, we have a very convenient
situation: the $W_8$-$\nu_8$ loops cancel the self-energy contribution of
the $Z_8$-$\mu_8$ loops, {\it but hardly affect the corresponding contribution
to $g-2$ of the muon}. This is a rare and satisfying situation in beyond-SM
physics and can ultimately be traced to the chiral structure of the
couplings\cite{Marciano} --- itself a consequence of the colour assignments
of the extra gauge bosons.

We have not exhibited a detailed analysis of the self-energy contributions
since these depend on several unknown parameters, viz., the masses of the
$W_8$ and $Z_8$ bosons, the masses of the coloured leptons $\mu_8$ and
$\nu_8$ (which have to be nearly degenerate to avoid constraints from
electroweak precision tests) and the magnitudes of the $W_8$-$\nu_8$-$\mu$
and $Z_8$-$\mu_8$-$\mu$ coupling constants. All that we require is a sufficiently
sensitive cancellation and we have checked that, even keeping the couplings and
boson masses of the same order in magnitude, enough leeway remains in the
parameter space to arrange for a phenomenological cancellation of the
self-energy corrections. It follows that the self-energy yields no useful 
constraints on the parameters responsible for the muon anomaly.

The contribution of a heavy fermion and a heavy neutral vector boson to
the anomalous magnetic moment of the muon has been evaluated several times
before\cite{Leveille}, and we skip the details of the actual derivation. The
one-loop-corrected vertex function of the muon can be written
\begin{equation}
\Gamma_\alpha = F_1(q^2) \gamma_\alpha
            + \frac{i}{2m_\mu} F_2(q^2) \sigma_{\alpha\beta}q^\beta
\end{equation}
from which it follows that
\begin{equation}
a_\mu = \frac{g -2}{2} = \left. F_2(q^2)\right|_{q^2 \to 0} \ .
\end{equation}
Direct evaluation\cite{Leveille} of the form factor $F_2(q^2)$ using the
Feynman rule of Fig.~1 and taking account of the colour factor of 8 leads
to the result (in terms of the unknown parameters $g_8, m_8, M_8$)
\begin{eqnarray}
a_\mu(Z_8) = 
\left( \frac{\alpha_8}{2\pi} \right) 
\left(\frac{m_\mu}{M_8} \right) \frac{\sqrt{x_8}}{(1 - x_8)^4}
&\bigg[& 4 - 7x_8 + 3x_8^2 - x_8^3 + x_8^4 + 6 x_8 (1 - x_8) \log x_8 
\nonumber \\
&& + \frac{1}{12}\frac{m_\mu}{M_8} \frac{1}{\sqrt{x_8}} 
\bigg( 8 - 38x_8 + 39x_8^2 - 14x_8^4 + 5 - 18x_8^2 \log x_8  \bigg) \bigg] \ ,
\nonumber 
\end{eqnarray}
where $\alpha_8 = g_8^2 /4\pi$ and $x_8 = m_8^2/M_8^2$. For large values 
of $M_8$ and $m_8 > 100$~GeV, the terms on the last line of the above 
equation may be safely neglected.

We are now in a position to make a phenomenological analysis of the haplon
model vis-\'a-vis the E821 data. Our numerical results are set out in
Fig.~2. In the left box we have plotted, for three different choices 
of $\alpha_8$ (= 0.1, 0.01, 0.001), the region in the $M_8$--$m_8$ plane 
which can explain the BNL 2.7$\sigma$ excess which arises using the estimate
based on $e^+e^-$ data. The darker-shaded regions correspond to the data
with errors at the 1$\sigma$ level, while the lighter-shaded regions
correspond to the 2$\sigma$ level. It is of great interest to note that
large values of the exotic particle masses $m_8$ and $M_8$ can actually
yield the correct order of magnitude of the anomalous magnetic
moment. This is because the actual contribution depends only on the ratio
of a function of $x_8$ with $M_8$.

\begin{figure}[h]
\begin{center}
\vspace*{3.0in}
      \relax\noindent\hskip -6.4in\relax{\includegraphics{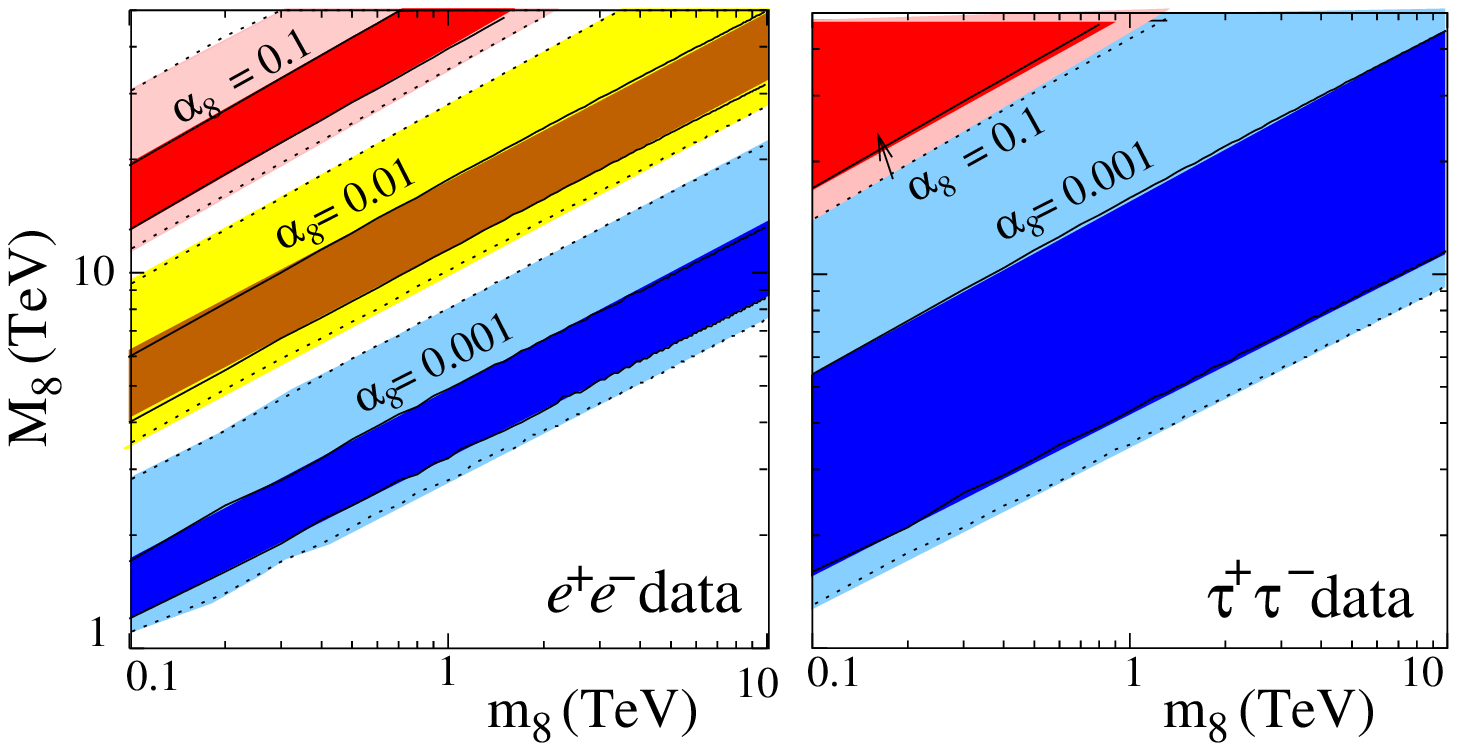}}
\end{center}
\end{figure}
\vspace*{-0.3in}
\noindent {\bf Figure 2}.
{\footnotesize\it Illustrating the regions in the $M_8 - m_8$ plane
which can explain the BNL data. Values of $\alpha_8$ are marked inside the
relevant shaded region. Dark (light) shading denotes the allowed region at
1(2)$\sigma$. There is no upper bound on $M_8$ at the 3$\sigma$ level
because the observed deviation from the SM is at 2.7$\sigma$. For the
$\tau^+\tau^-$ data, almost all large values of $M_8, m_8$ are allowed; 
hence the choice $\alpha_8 = 0.01$ has not been exhibited.} 
\vskip 5pt

We see, therefore, that it is possible to interpret the observed deviation
of the muon anomalous moment from the SM prediction as a signal for a
preonic structure of the weak gauge bosons provided the effective coupling
$\alpha_8$ lies in the range 0.001 to 0.1 and other parameters of the
model guarantee sufficiently small corrections to the muon mass. The lower
value of $\alpha_8$ represents a superweak-type interaction, while the
upper value represents a strong interaction. However, it is gratifying to
note that the most favourable region in the parameter space is indeed when
$\alpha_8$ is close to the weak coupling strength, {\it i.e.} ${\cal
O}(0.01)$. This value is an eminently desirable feature, given the
philosophy adopted in the haplon model. In fact, several interesting
phenomenological features of the model emerge. For example, the exotic
particle masses which can explain the BNL result are rather large and can
easily evade any constraints coming from Fermilab Tevatron data. In fact,
to see any manifestation of these exotic particles and their interactions
at low energies will generically require a measurement which is precise to
the same level as the E821 experiment, which would explain why the
observed excess in the muon anomalous magnetic moment could be a single
isolated hint of a composite structure of leptons and weak bosons. However,
we must note that the coloured $Z$ and coloured muon states cannot be
decoupled, i.e. the masses cannot be arbitrarily heavy. Thus, a future
experiment which would scan the 10--50~TeV range might actually produce
these particles directly.

As we have noted before, it is also necessary to take into account the 
possibility that the correct theoretical estimate is the one obtained from
the $\tau^\pm$ data and hence that the BNL measurement is more or less 
consistent with the SM prediction. In this case, the measurement can only 
be used to constrain the haplon model. We have plotted the allowed region in
Fig.~2 in the box on the right, for two extreme values $\alpha_8$ (= 0.1, 
0.001). As expected, the experimental result forces the $\mu_8$ and the
$Z_8$ to be rather heavy --- in fact, the $Z_8$ must be over 1~TeV even 
when the coupling $\alpha_8$ is rather small. However, once this is taken,
most of the parameter space is allowed, which is consistent with a decoupling
behaviour, and is obviously, less exciting. 

To summarize, then, we have made a phenomenological analysis of the recent
BNL measurement and shown how the possible 2.7$\sigma$ excess can be 
interpreted as a signal for heavy
exotic coloured lepton and $Z$-boson states arising in a preonic model.
Though, admittedly, this is not a unique way in which the observed
deviation from the SM can be interpreted, nor is the haplon model the only
preonic model, but our result is nevertheless exciting, since it means
that the old idea of substructure remains one of the new physics options
which remain viable in the post-BNL scenario. This does not depend very
crucially on our choice of model (the haplon model was essentially chosen for
simplicity) but it requires a degree of fine-tuning in the muon
self-energy whatever be the model chosen. Whether this explanation is the
correct one, or whether the postulated substructure is anywhere near the
correct theory is something only time and new data from other processes
can indicate.

\bigskip\bigskip

\noindent {\bf Acknowledgments}: The authors collectively thank Pankaj
Jain, the late Uma Mahanta and Gautam Sengupta for discussions. 


\end{document}